\documentclass{llncs}

\usepackage{amssymb,amsmath,graphicx}
\usepackage{pgf}
\usepackage{color,enumerate,comment,boxedminipage}
\usepackage{float}
\usepackage{hyperref}

\newcommand{\problemdef}[3]{
	\begin{center}
		\begin{boxedminipage}{.99\textwidth}
			\textsc{{#1}}\\[2pt]
			\begin{tabular}{ r p{0.8\textwidth}}
				\textit{~~~~Instance:} & {#2}\\
				\textit{Question:} & {#3}
			\end{tabular}
		\end{boxedminipage}
	\end{center}
}

\newcommand{\ftrans}{$\mathcal{F}$-transversal}

\newcommand{\NP}{{\sf NP}}

\title{Minimum Connected Transversals in Graphs:\\ New Hardness Results and Tractable Cases\\ Using the Price of Connectivity\thanks{This work was supported by a London Mathematical Society Scheme $4$ Grant, Leverhulme Trust Grant RPG-2016-258 and by the Slovenian Research Agency (I$0$-$0035$, research programs P$1$-$0285$, research projects N$1$-$0032$, J$1$-$5433$, J$1$-$6720$, J$1$-$6743$, J$1$-$7051$, and a Young Researchers Grant).}}

\author{
Nina Chiarelli \inst{1}
\and
Tatiana R. Hartinger \inst{1}
\and
Matthew Johnson\inst{2}
\and\\
Martin Milani\v{c}\inst{1}
\and
Dani\"el Paulusma \inst{2}
}

\institute{
University of Primorska, UP IAM and UP FAMNIT, Koper, Slovenia\\
\texttt{nina.chiarelli@famnit.upr.si,tatiana.hartinger@iam.upr.si,martin.milanic@upr.si}
\and
School of Engineering and Computing Sciences, Durham University, UK\\
\texttt{\{matthew.johnson2,daniel.paulusma\}@durham.ac.uk}
}

\begin{document}

\maketitle

\begin{abstract}
We perform a systematic study in the computational complexity of the connected variant of three related transversal
problems: {\sc Vertex Cover}, {\sc Feedback Vertex Set}, and {\sc Odd Cycle Transversal}. Just like their original counterparts, these variants are \NP-complete for general graphs. A graph~$G$ is $H$-free for some graph $H$ if $G$ contains no induced subgraph isomorphic to~$H$.
It is known that {\sc Connected Vertex Cover} is \NP-complete even for $H$-free graphs if $H$ contains a claw or a cycle.
We show that the two other connected variants also remain \NP-complete if $H$ contains a cycle or claw. In the remaining case $H$ is a linear forest. We show that {\sc Connected Vertex Cover}, {\sc Connected Feedback Vertex Set}, and {\sc Connected Odd Cycle Transversal} are polynomial-time solvable for $sP_2$-free graphs for every constant $s\geq 1$. For proving these results we use known results on the price of connectivity for
vertex cover, feedback vertex set, and odd cycle transversal. This is the first application of the price of connectivity that results in polynomial-time algorithms.
\end{abstract}

\section{Introduction}\label{s-intro}

We consider graph modification problems, where the aim is to test if a given graph can be modified to have some desired property using a small number of graph operations.  Many such problems can be formulated as a vertex transversal problem.
That is, the modified graph, obtained after deleting a number of vertices of the original graph, must not contain any graph of a prescribed family~${\cal F}$. More formally, we are looking for a small \ftrans\ of a graph $G=(V,E)$, where an {\it \ftrans} of $G$ is a subset $S \subseteq V$ such that $G-S$ is {\it ${\mathcal F}$-free}. In other words, $S$ intersects every subset of $V$ that induces a subgraph isomorphic to a graph in $\mathcal{F}$.
For instance, let 
$P_k$ denote the path on $k$ vertices.
Then a set $S$ is a \emph{vertex cover} if $S$ is a  $\{P_2\}$-transversal. Note that $V-S$ is an independent set. A set $S$ is a \emph{feedback vertex set} if $S$ is an \ftrans\ for the infinite family
${\cal F}=\{C_3,C_4,C_5,\ldots\}$
(where $C_k$ is the cycle on $k$ vertices). In this case, the graph $G-S$ is a forest.
To give another example, a set $S$ is an {\it odd cycle transversal} if $S$ is an \ftrans\ for the infinite family
${\cal F}=\{C_3,C_5,C_7,\ldots\}$. In this case, the graph $G-S$ is bipartite.

Vertex covers, feedback vertex sets, and odd cycle transversals are well-studied concepts.
The corresponding decision problems that ask whether a given graph has a vertex cover, feedback vertex set, or
odd cycle transversal of size at most~$k$ for some given integer~$k$ are called {\sc Vertex Cover},
{\sc Feedback Vertex Set} and {\sc Odd Cycle Transversal}, respectively.
These problems are well known to be \NP-complete~\cite{GJ79,LY80}.
In this paper we focus on the connected variants of these problems. We say that a vertex cover $S$ of a graph $G$ is connected if $S$ induces a connected subgraph of $G$. The notions of a connected feedback vertex set and connected odd cycle transversal are defined in a similar way.

\problemdef{{\sc Connected Vertex Cover}}{a graph $G$ and an integer $k$.}{does $G$ have a connected vertex cover $S$ with $|S|\leq k$?}

\problemdef{{\sc Connected Feedback Vertex Set}}{a graph $G$ and an integer $k$.}{does $G$ have a connected feedback vertex set $S$ with $|S|\leq k$?}

\problemdef{{\sc Connected Odd Cycle Transversal}}{a graph $G$ and an integer $k$.}{does $G$ have a connected odd cycle transversal $S$ with $|S|\leq k$?}

\subsection{Related Work}

The three connected problem variants are known to be  \NP-complete. It is therefore a natural question whether there exist special graph classes for which they become polynomial-time solvable. Below we briefly survey known hard and tractable results for each of the three problems.

Already in the seventies, Garey and Johnson~\cite{GJ77} proved that  {\sc Connected Vertex Cover} is \NP-complete even for planar graphs of maximum degree~4. Later, Priyadarsini and Hemalatha~\cite{PH08} strengthened this result to
2-connected planar graphs of maximum degree~4, and Fernau and Manlove~\cite{FM09} strengthened it to planar bipartite
graphs of maximum degree~4. Wanatabe, Kajita, and Onaga~\cite{WKO91} proved \NP-completeness of {\sc Connected Vertex Cover} for 3-connected graphs.
More recently, Munaro~\cite{Mu} established \NP-completeness of {\sc Connected Vertex Cover} for line graphs
of planar cubic bipartite graphs
and for
planar bipartite
graphs of arbitrarily large girth.
Ueno, Kajitani, and Gotoh~\cite{UKG88} proved that {\sc Connected Vertex Cover} can be solved in polynomial time for graphs of maximum degree at most~3 and for trees.
Escoffier, Gourv\`es, and Monnot~\cite{EGM10} improved the latter result by showing that {\sc Connected Vertex Cover} is
polynomial-time solvable for chordal graphs. It is known that {\sc Vertex Cover} is polynomial-time solvable for chordal graphs as well. The same authors proposed to study
the complexity of {\sc Connected Vertex Cover} for
other graph classes for which {\sc Vertex Cover} is polynomial-time solvable.

Grigoriev and Sitters~\cite{GS09} proved that {\sc Connected Feedback Vertex Set} is \NP-complete for planar graphs with maximum degree~9. Besides this result not much is known on the computational complexity of this problem except that it is fixed parameter tractable when parameterized by~$k$~\cite{MPRSS12}.  The {\sc Connected Odd Cycle Transversal} has been mainly studied from
parameterized point of view~\cite{CNPPRW11,CPPW10}.
It is implicit in these works that {\sc Connected Odd Cycle Transversal} is \NP-complete, though we were not able to find any proof of this (simple) result.  For the sake of completeness, we note now that it is implied by the stronger results we present in Theorems~\ref{t-coct} and~\ref{t-coct2}.

\subsection{Our Results.}
We continue the investigations on the complexity of {\sc Connected Vertex Cover} and also initiate a complexity study of {\sc Connected Feedback Vertex Set} and
{\sc Connected Odd Cycle Transversal}. Our paper consists of two parts. In the first part (Section~\ref{s-hard}) we prove new
hardness results and in the second part (Section~\ref{s-easy}) new tractable results. As we explain below we try to do so
in a systematic way. We refer to Table~\ref{t-thetable} for a survey of our results together with some related known results. This table leads to some natural open problems, which we discuss in Section~\ref{s-con}, together with some other directions for future work.

The {\it girth} of a graph is the length of a shortest cycle in it. We prove that
{\sc Connected Feedback Vertex Set} and {\sc Connected Odd Cycle Transversal} are \NP-complete on graphs of
girth at least~$p$ for any fixed constant~$p\geq 3$. In order to obtain this result for {\sc Connected Odd Cycle Transversal}
we first prove \NP-completeness of {\sc Odd Cycle Transversal} for graphs of girth at least $p$ for any fixed constant
$p\geq 3$ (analogous results were already known for {\sc Vertex Cover} and {\sc Feedback Vertex Set}).
In the same section we also show that {\sc Connected Feedback Vertex Set} and {\sc Connected Odd Cycle Transversal} are
\NP-complete for line graphs.

The above hardness results imply that
{\sc Connected Feedback Vertex Set} and {\sc Connected Odd Cycle Transversal} are
\NP-complete on $H$-free graphs whenever $H$ is a graph that contains a cycle or a claw.
Due to the results of Munaro~\cite{Mu} the same holds for {\sc Connected Vertex Cover}.
Moreover, the same holds for the original variants
{\sc Feedback Vertex Set} and {\sc Odd Cycle Transversal} but not for {\sc Vertex Cover}, which is polynomial-time solvable for claw-free graphs.
Due to the hardness results we can restrict ourselves to $H$-free graphs where $H$ is a {\it linear forest}, that is, the disjoint union of one or more paths.

We prove that {\sc Connected Vertex Cover}, {\sc Connected Feedback Vertex Set}, and {\sc Connected Odd Cycle Transversal} are polynomial-time solvable for $sP_2$-free graphs for every constant $s\geq 1$.
Since {\sc Vertex Cover} is polynomial-time solvable for $sP_2$-free graphs for any $s\geq 1$,\footnote{This follows from combining results of \cite{BY89,TIAS77}, as explained in more detail in Section~\ref{s-easy}.}
our result for {\sc Connected Vertex Cover} restricted to $sP_2$-free graphs can be seen as a new step in the aforementioned research direction of
Escoffier, Gourv\`es, and Monnot~\cite{EGM10}.

\begin{table}[h]
\centering
\begin{tabular}{|l|c|l|l|c|l|}
\hline
& girth~$p$ & line graphs & $sP_2$-free\\[-1pt]
\hline
{\sc Con. Vertex Cover}  &\hspace*{6mm}\NP-c~\cite{Mu} &\NP-c~\cite{Mu} &P\\[-1pt]
\hline
{\sc Con. Feedback Vertex Set} &\NP-c &\NP-c &P\\[-1pt]
\hline
{\sc  Con. Odd Cycle Transversal} &\NP-c &\NP-c &P\\[-1pt]
\hline
{\sc Vertex Cover} &\hspace*{6mm} \NP-c~\cite{Po74}\; &P\hspace*{4.7mm} \cite{Sh80}&P~\cite{BY89,TIAS77}\\[-1pt]
\hline
{\sc Feedback Vertex Set} &\hspace*{6mm} \NP-c~\cite{Po74}\;  &\NP-c~\cite{Sp83} &P\\[-1pt]	
\hline
{\sc Odd Cycle Transversal} &\NP-c &\NP-c &P\\[-1pt]	
\hline			
\end{tabular}
\vspace*{2mm}
\caption{The complexities of the three connected transversal problems together and the original transversal problems for graphs of girth at least~$p$ for every constant $p\geq 3$, line graphs,
or $sP_2$-free graphs for any $s\geq 1$. The results for {\sc Vertex Cover} and {\sc Feedback Vertex Set} on graphs of arbitrarily large girth follow from Poljak's construction~\cite{Po74}
(see also Section~\ref{s-hard}).
All unreferenced results are new results from this paper.}\label{t-thetable}
\end{table}

To prove our tractability results we first focus on the original problems {\sc Vertex Cover}, {\sc Feedback Vertex Set}, and
{\sc Odd Cycle Transversal}. In particular we will use the enumeration algorithms of Tsukiyama, Ide, Ariyoshi and Shirakawa~\cite{TIAS77} and Schwikowski and Speckenmeyer~\cite{SS02} for listing all maximal independent sets and all maximal feedback vertex sets, respectively. We will also use the fact that the number of maximal independent sets is polynomially bounded for $sP_2$-free graphs~\cite{BY89}. However, we need to prove analogous results for
feedback vertex sets and odd cycle transversals. This leads to polynomial-time enumeration algorithms for minimal feedback vertex sets of an $sP_2$-free graph and minimal odd cycle transversals of an $sP_2$-free graph. What remains to do is to relate these results to the connected problem variants and below we describe how to do this.

We need to consider the effect of adding the connectivity constraint on the minimum size of an ${\cal F}$-transversal for a graph family ${\cal F}$. This effect is in fact measured by a known concept called the {\it price of connectivity}.
This concept was coined by Cardinal and Levy~\cite{CL10} for vertex cover, but can be defined
for any graph property for which a connected variant is meaningful:
for a class of graphs ${\cal G}$ and graph property~$\pi$, the price of connectivity is the
worst-case ratio $\pi'(G)/\pi(G)$ over all connected graphs $G\in {\cal G}$, where $\pi(G)$ and $\pi'(G)$ denote the smallest subset and smallest connected subset, respectively, of the vertices of $G$ satisfying $\pi$.
Apart from further results on vertex cover~\cite{Ca,CC17,CCFS14}, the price of connectivity has been studied for
dominating set~\cite{Ca,CC17,CS14,DM82,Zve03}, face hitting set~\cite{GS09,SS10}, and feedback vertex set~\cite{BHKP17}, until
in~\cite{HJMP16} known results for vertex cover and feedback vertex set for $H$-free graphs were extended in a larger framework
of ${\cal F}$-transversals. In particular, it is known that for $sP_2$-free graphs demanding connectivity results only in an additive constant with respect to the size of a smallest vertex cover, feedback vertex set, or odd cycle transversal~\cite{BHKP17,HJMP16}. This is exactly what we need to finalize our proofs.

\section{Hardness Results}\label{s-hard}

In this section we prove our hardness results (see also Table~\ref{t-thetable}). We first need to introduce some terminology.
The {\em line graph}~$L(G)$ of a graph $G=(V,E)$ has the edges of~$E$ as its vertices, and two vertices~$e_1$ and~$e_2$ of~$L(G)$ are adjacent if and only if~$e_1$ and~$e_2$ have an end-vertex in common in~$G$.
The {\em claw} is the graph with vertices~$s$, $t_1$, $t_2$, $t_3$ and edges~$st_1$, $st_2$,~$st_3$.
It is well-known and readily seen that every line graph is claw-free.

Our proofs for the
{\sc Odd Cycle Transversal} and {\sc Connected Odd Cycle Transversal} problems can be found in Section~\ref{s-3} and for the {\sc Connected Feedback Vertex Set} problem in Section~\ref{s-2}.

\subsection{Odd Cycle Transversal and Connected Odd Cycle Transversal}\label{s-3}
We will first prove that {\sc Odd Cycle Transversal} and {\sc Connected Odd Cycle Transversal} are \NP-complete for line graphs.

A {\it 2-factor} of a graph $G$ is a set of cycles such that every vertex of $G$ belongs to exactly one of them.
Let {\sc Even Cycle Factor} denote the problem of deciding if a graph has an {\it even cycle factor}, that is, a 2-factor in which each cycle is even.

We need the following lemmas due to Hell, Kirkpatrick, Kratchov\'il, and K\v{r}\'i\v{z}.

\begin{lemma}[\cite{HKKK88}]\label{l-twofactor}
{\sc Even Cycle Factor} is \NP-complete.
\end{lemma}

We also need the next lemma. Here, we let $G[S]$ denote the subgraph of a graph $G$ induced by a subset $S\subseteq V(G)$.

\begin{lemma} \label{l-line}
Let $G$ be a graph on $n$ vertices and $m$ edges.  If $L(G)$ has an odd cycle transversal of size at most $m-n$, then $G$ has an even cycle factor.
\end{lemma}

\begin{proof}
Let $T$ be the vertices of $L(G)$ that do not belong to the odd cycle transversal.  As $L(G)$ has $m$ vertices, $T$ contains at least $n$ vertices.  By definition,
$L(G)[T]$ is bipartite which means in particular that the neighbours of each vertex are independent.  Thus,
as $L(G)$ is a line graph and so claw-free, no vertex in $L(G)[T]$ has degree greater than~2.  Hence $L(G)[T]$ is the disjoint union of a set of paths and even cycles.   Each path on $r$ vertices in $L(G)$ corresponds to a path on $r+1$ vertices in $G$, and each cycle on $r$ vertices in $L(G)$ corresponds to a cycle on $r$ vertices in $G$.  Thus, if $p$ of the components of $G[T]$ are paths, then
the number of vertices of $G$ incident with $T$ is $|T|+p$.
As $T$ has at least $n$ vertices and $G$ has exactly $n$ vertices, we must have $p=0$, and so $T$ is the disjoint union of even cycles that corresponds to an even cycle factor of $G$.
\qed
\end{proof}

\begin{theorem}\label{t-oct}
{\sc Odd Cycle Transversal} is \NP-complete for line graphs.
\end{theorem}

\begin{proof}
We reduce from {\sc Even Cycle Factor}, which is \NP-complete by Lemma~\ref{l-twofactor}. Let $G$ be an instance of this problem with $n$ vertices and $m$ edges.
We claim that $G$ has an even cycle factor if and only if its line graph $L(G)$ has an odd cycle transversal of size at most~$m-n$.

If $G$ has an even cycle factor, then $L(G)$ contains a set $T$ of $n$ vertices that induce a disjoint union of cycles. Hence $S=V(L(G))\setminus T$ is an odd cycle transversal of $L(G)$ that has size $|L(G)|-n=m-n$.  To complete the proof, we apply Lemma~\ref{l-line}.
\qed
\end{proof}

\begin{theorem}\label{t-coct}
{\sc Connected Odd Cycle Transversal} is \NP-complete for line graphs.
\end{theorem}

\begin{proof}
As in the proof of Theorem~\ref{t-oct} we reduce from {\sc Even Cycle Factor}. Let $G$ be an instance of this problem with $n$ vertices and $m$ edges. We add a new vertex $x$ to $G$ and an edge between $x$ and each vertex of $G$. We also add three vertices $y_1,y_2,y_3$ with edges $xy_1$, $y_1y_2$, $y_2y_3$ and $y_3x$.
We call the resulting graph $G'$. We observe that $G$ has an even cycle factor if and only if $G'$ has an even cycle factor. Note that $G'$ has $n+4$ vertices and $m+n+4$ edges.
We construct the line graph $L(G')$ of $G'$ and
claim that $G'$ has an even cycle factor if and only if $L(G')$ has a connected odd cycle transversal of size at most~$m$.

First suppose that $G'$ has an even cycle factor $F$. By construction, $F$ contains the cycle on vertices $x,y_1,y_2,y_3$, which we denote by $C$. As a consequence, no edge $xu$ for $u\in V(G)$ belongs to a cycle of $F$.
Then $L(G')$ contains a set $T$ of $n+4$ vertices that induce a disjoint union of cycles that do not contain any vertex $xu$
for $u\in V(G)$.
Hence $S=V(L(G'))\setminus T$ is an odd cycle transversal of $L(G')$ that has size
$|L(G')|-|V(F)|=m+n+4-(n+4)=m$ and that contains all vertices of the form $xu$. We need to show that $S$ induces a connected subgraph of $L(G')$.
This follows from the fact that every vertex
$e=uv\in S\cap E(G)$ is adjacent to $xu$ (and $xv$), which belongs to $S$, and the
fact that the set $\{xu\; |\; u\in V(G)\}$ form a clique in $L(G')$.

Now suppose that  $L(G')$ has a connected odd cycle transversal of size at most~$m$. Then we can apply Lemma~\ref{l-line} (noting again that $G'$ has $n+4$ vertices and $m+n+4$ edges) to find that $G'$ has an even cycle factor.
\qed
\end{proof}

For our next hardness result on {\sc Connected Odd Cycle Transversal} we first need to show the same result
for {\sc Odd Cycle Transversal}. For that we need the following lemma.

\begin{lemma}\label{l-twice}
Let $G'$ be a graph obtained from a graph $G$ after subdividing an edge of $G$ twice.
Then the size of  a minimum odd cycle transversal of $G$ is equal to the size of a minimum odd cycle transversal of $G'$.
\end{lemma}

\begin{proof}
Let $uv$ be the edge of $G$ that is subdivided twice. Let $a,b$ be the two new vertices obtained after the subdividing so that $G'$ contains the path $uabv$.

First suppose that $S$ is a minimum odd cycle transversal of $G$. If one or two of $u,v$ are in $S$, then $S$ is an odd cycle transversal of $G'$ (as we can safely place $a,b$ in $G'-S$). The same is true if
$u,v\notin S$, as any even cycle in $G-S$ is still even after subdividing $e$ twice.

Now suppose that $S'$ is a minimum odd cycle transversal of $G'$. If one or two of $u,v$ are in $S'$, then $a,b$ are not in $S'$ by minimality of $S'$, meaning that $S'$ is also an odd cycle transversal of $G$. Suppose that $u,v$ are not in $S'$. If $a$ or $b$ is in $S'$, then we can safely replace $a$ or $b$ by $u$ or $v$, respectively, to obtain a new odd cycle transversal of $G'$ with the same size as $S'$.
If none of $u,a,b,v$ is in $G'-S'$, then, as contracting the path
$uabv$ back to $uv$ does not result in an odd cycle, we again find that $S'$ is an odd cycle transversal of $G$.
\qed
\end{proof}

Repeatedly applying Lemma~\ref{l-twice} in combination with the fact that {\sc Odd Cycle Transversal} is \NP-complete (due to Theorem~\ref{t-oct}) implies the following theorem.

\begin{theorem}\label{t-oct2}
For every constant $p\geq 3$, {\sc Odd Cycle Transversal}  is \NP-complete for graphs of girth at least~$p$.
\end{theorem}

We now show that we also have hardness when the problem asks for a connected transversal.

\begin{theorem}\label{t-coct2}
For every constant $p\geq 3$, {\sc Connected Odd Cycle Transversal} is \NP-complete for graphs of girth at least~$p$.
\end{theorem}

\begin{proof}
We reduce from {\sc Connected Vertex Cover} restricted to graphs of girth at least~$p$.
Recall that Munaro~\cite{Mu} proved that this problem is \NP-complete.
 Let $G$ be a graph of girth at least~$p$.
 We may assume without loss of generality that $G$ has at least two edges.
 Between each pair of adjacent vertices of $G$ add a path with $2 \lfloor \frac{p}{2} \rfloor$ edges.  Let the graph obtained be $G'$ and note that it also has girth at least~$p$.  We will show that $G$ has a connected vertex cover of size at most $k$ if and only if $G'$ has a connected odd cycle transversal of size at most $k$.

Let $S$ be a connected vertex cover of $G$ of at most $k$.  As every cycle in $G'$ contains two vertices that are adjacent in $G$,
the set $S$ is also an odd cycle transversal of $G'$, and it is also connected in $G'$ as $G'$ contains all the edges of $G$.

Let $S$ be a connected odd cycle transversal of $G'$ of at most $k$.  We can assume that all vertices of $S$ belong to $G$
(if this is not the case, then we can move all vertices from $S\setminus V(G)$ to $G'-S$ without creating an odd cycle in $G'-S$ or
disconnecting $G'[S]$).  As each pair of adjacent vertices in $G$ belong to the odd cycle of $G'$ that contains them and the path on $2 \lfloor \frac{p}{2} \rfloor$ edges added between, at least one of them must belong to $S$.  Thus $S$ is a connected vertex cover of $G$ (of size at most~$k$).
\qed
\end{proof}

Theorems~\ref{t-coct} and~\ref{t-coct2} imply the following corollary.

\begin{corollary}\label{c-coct}
Let $H$ be a graph. Then
{\sc Connected Odd Cycle Transversal} is \NP-complete for $H$-free graphs if $H$ contains a cycle or a claw.
\end{corollary}

\subsection{Connected Feedback Vertex Set}\label{s-2}

We will now consider the {\sc Connected Feedback Vertex Set} problem.
We first prove the following result.

\begin{theorem}\label{t-cfvs}
{\sc Connected Feedback Vertex Set} is \NP-complete for line graphs.
\end{theorem}

\begin{proof}
Let $G$ be an instance of {\sc Hamiltonian Path} with $n\geq 3$ vertices and $m$ edges.
We will describe a reduction to {\sc Connected Feedback Vertex Set}. We add two new vertices $x$ and $y$ to $G$ and an edge between $x$ and each vertex of $G$ and an edge between $y$ and each vertex of $G$.
We also add two vertices $x'$ and $y'$ and edges $x'x$ and $y'y$.
We call the resulting graph $G'$. We observe that $G$ has a Hamiltonian path if and only if $G'$ has a Hamiltonian path
from $x'$ to $y'$.
Note that $G'$ has $n+4$ vertices and $m+2n+2$ edges.
We
claim that $G'$ has a Hamiltonian path if and only if its line graph $L(G')$ has a connected feedback vertex set of size at most
$m+n-1$.

First suppose that $G'$ has a Hamiltonian path $P$ (on $n+4$ vertices). Let $xa$ and $yb$ be the two edges of $P$
that join $x$ and $y$ to $V(G)$.
Then every edge $xu$ with $u\in V(G)\setminus \{a\}$  and every edge $yv$ with $v\in V(G)\setminus \{b\}$
does not belong to $P$.
Let $T \subseteq V(L(G'))$ be the set of size $n+3$
in $G'$ formed by the edges of $P$.
Hence $S=V(L(G'))\setminus T$ is a feedback vertex set of $L(G')$ that has size
$m+2n+2-(n+3)=m+n-1$ and that contains all vertices of the form $xu$ and $yu$ except the vertices $xa$ and
$yb$. We need to show that $S$ induces a connected subgraph of $L(G')$.
This follows from the fact that every vertex $e=uv\in S\cap E(G)$ is adjacent to $xu$ and $yu$, and at least one of them belongs to $S$. Moreover, the set of vertices $\{xu\; |\; u\in V(G)\setminus \{a\}\}$ is a clique in $L(G')$. Finally, every vertex $yu\in S$ with
$u\in V(G)\setminus \{a,b\}$ is adjacent to $xu$ in $L(G')$.

Now suppose that $L(G')$ has a connected feedback vertex set $S$ of size at most $m+n-1$. Then $L(G')-S$ has
at least $n+3$ vertices and must be a disjoint union of paths (as $L(G')$ is a line graph and thus claw-free).
Every induced path in $L(G')$ corresponds to a path in $G'$ with one more vertex. Moreover, $G'$ has $n+4$ vertices in total. Hence $L(G')-S$ must induce a single path on $n+3$ vertices. The corresponding  path in $G'$ has $n+4$ vertices and is thus a Hamiltonian path of~$G'$.\qed
\end{proof}

We now show that the restriction of {\sc Connected Feedback Vertex Set} to graphs of arbitrarily large girth~$p$ is \NP-hard.
We note that it is known that {\sc Feedback Vertex Set} is \NP-complete for graphs of girth at least~$p$ for every constant~$p\geq 3$; this follows immediately from the well-known observation (see for example~\cite{BDFJP,MPRS12}) that if a graph $G'$ is obtained from a graph $G$ by subdividing an edge, then $G$ has a feedback vertex set of size at most~$k$ if and only if~$G'$ does. In order to prove the result for {\sc Connected Feedback Vertex Set}, we reduce from {\sc Connected Vertex Cover} 
restricted to graphs of girth at least $p$ using essentially the same construction and arguments as in the proof of Theorem~\ref{t-coct2}. In this case, we can simply say that for an instance $G,k$ of {\sc Connected Vertex Cover}, we construct an instance $G',k$ of {\sc Connected Feedback Vertex Set} by adding a path of length $p-1$ between each pair of adjacent vertices in $G$ (so that each edge of $G$ corresponds to a cycle of length~$p$ in $G'$).

\begin{theorem}\label{t-cfvs2}
For every constant $p\geq 3$, {\sc Connected Feedback Vertex Set} is \NP-complete for graphs of girth at least~$p$.
\end{theorem}

Theorems~\ref{t-cfvs} and~\ref{t-cfvs2} imply the following corollary.

\begin{corollary}\label{c-cfvs}
Let $H$ be a graph. Then
{\sc Connected Feedback Vertex Set} is \NP-complete for $H$-free graphs if $H$ contains a cycle or a claw.
\end{corollary}

\section{Polynomial-Time Results}\label{s-easy}

In this section we prove our polynomial-time results (see also Table~\ref{t-thetable}).
The following lemma provides a useful technique that we will be able to apply several times.  We say that a set of vertices in a graph is \emph{connected} if it induces a connected subgraph.

\begin{lemma} \label{l-polyfind}
Let $P$ be a property of a set of vertices in a graph
that if it holds for a set also holds for all supersets.  Given a graph~$G$, let $f(G)$ be the minimum size of a set of vertices in $G$ with property $P$. Let $\cal G$ be a graph class and let $c$ be a constant.  Suppose that for every graph $G \in {\cal G}$, all minimal sets with property $P$ can be enumerated in polynomial time, and the minimum size of a connected set of vertices with property $P$ is at most $f(G)+c$. Then a minimum size connected set of vertices of $G$ with property $P$ can be found in polynomial time.
\end{lemma}

\begin{proof}
The algorithm for finding a minimum size connected set with property $P$ is simple. Enumerate the minimal sets of $G$ with property $P$, and, for each such set consider all the sets that can be obtained by adding $c$ or fewer additional vertices and check whether each is connected. Return the smallest connected set found (if any).  It is clear that the algorithm runs in polynomial time.  We must prove correctness.

Let $S$ be a minimum size connected set of vertices in $G$ with property $P$.  We will show that the algorithm finds $S$ and so returns it (or another connected set with property~$P$ of the same size).  We know that $|S| \leq f(G)+c$.  From $S$ remove as many vertices as possible until a minimal set $S'$ with property $P$ is obtained.   Then $S'$ is one of the sets that will be enumerated and as $|S'| \geq f(G)$, we have that $|S|-|S'|\leq c$ and so $S$ will be found.
\qed
\end{proof}

For an algorithm that enumerates all solutions to a problem, the \emph{delay} is the maximum of the time taken before finding the first solution and that required between any pair of consecutive outputs.
We need two well-known results. The first one is due to Balas and Yu (who gave a sharper bound, which we do not need for our purposes) and the second one is due to Tsukiyama, Ide, Ariyoshi, and Shirakawa.

\begin{theorem}[\cite{BY89}]\label{t-by}
For every constant~$s\geq 1$, the number of maximal independent sets of an $sP_2$-free graph on $n$ vertices is at most $n^{2s}+1$.
\end{theorem}

\begin{theorem}[\cite{TIAS77}]\label{t-tias}
For every constant~$s\geq 1$, it is possible to enumerate all maximal independent sets of a graph $G$ on $n$ vertices and $m$ edges
with a delay of $O(nm)$.
\end{theorem}

It is well-known that {\sc Vertex Cover} is polynomial-time solvable for $sP_2$-free graphs; in fact this follows immediately
from Theorems~\ref{t-by} and~\ref{t-tias}. Our proof of the same result for {\sc Connected Vertex Cover} can be found in Section~\ref{s-v}, for {\sc Feedback Vertex Set} and {\sc Connected Feedback Vertex Set} in Section~\ref{s-f}, and  for {\sc Odd Cycle Transversal} and {\sc Connected Odd Cycle Transversal} in Section~\ref{s-o}.

\subsection{Connected Vertex Cover}\label{s-v}

We consider the {\sc Connected Vertex Cover} problem.
We have the following result on the price of connectivity for
vertex cover
restricted to $sP_3$-free graphs by Lemma~10 of a previous paper of four of
the current authors.
The constant $4s^2+2s-10$ is determined in the proof of Lemma~10 of~\cite{BHKP17} (the statement of the lemma only mentions the existence of a constant without further specification).

\begin{theorem}[\cite{HJMP16}]\label{t-price2}
Let $s\geq 1$ and let $G$ be a connected $sP_3$-free graph. Let $f$ be the size of a minimum vertex cover of
$G$. Then the size of a minimum connected vertex cover of $G$ is at most $f+4s^2+2s-10$.
\end{theorem}

We can use Theorem~\ref{t-price2} to determine the complexity of {\sc Connected Vertex Cover} for $sP_2$-free graphs.

\begin{theorem}\label{t-main2}
For every constant~$s\geq 1$, {\sc Connected Vertex Cover} can be solved in polynomial time for $sP_2$-free graphs.
\end{theorem}

\begin{proof}
We know by Theorems~\ref{t-by} and~\ref{t-tias} that all maximal independent sets, and thus all minimal vertex covers, of $sP_2$-free graphs can be enumerated in polynomial time.  Thus, as the property of being a vertex cover holds for supersets, the theorem follows from Theorem~\ref{t-price2} and Lemma~\ref{l-polyfind} (note that we may restrict ourselves to connected input graphs). \qed
\end{proof}

\subsection{Feedback Vertex Set and Connected Feedback Vertex Set}\label{s-f}

To prove our results for {\sc Feedback Vertex Set } and {\sc Connected Feedback Vertex Set} we need the following result of
Schwikowski and Speckenmeyer~\cite{SS02}.

\begin{theorem}[\cite{SS02}]\label{t-ss}
It is possible to enumerate all minimal feedback vertex sets of a graph $G$ on $n$ vertices and $m$ edges
with a
delay of $O(n^3+n^2m)$.
\end{theorem}

We also need the following lemma.

\begin{lemma}\label{l-1}
For every constant~$s\geq 1$ there is a constant $c_s$ such that the number of minimal feedback vertex sets of an $sP_2$-free graph $G$ on $n$ vertices is
 $O(n^{c_s})$.
\end{lemma}

\begin{proof}
Let $S$ be a minimal feedback vertex set of $G$.
Let $F_S=G-S$, which, by definition, is a forest.
Let $F_S' \subset F_S$ contain each vertex of degree at least 2 in $F_S$ plus one vertex (arbitrarily chosen) from each component of $F_S$ isomorphic to $P_2$ (if there are any). We say that $F_S'$ is a {\it skeleton} of~$F_S$.
Let $\ell(F_S')=V(F_S)\setminus V(F_S')$ and notice that $\ell(F_S')$ is an independent set and each of its vertices has at most one neighbour in $F_S'$.  We say that $\ell(F_S')$ is the {\it leave} of $F_S'$.

\begin{claim}
$|V(F_S')| \leq 3s^2-5s+2$.
\end{claim}
To prove the claim, first consider the following subset $A$ of $V(F_S')$: all isolated vertices of $F_S'$, one arbitrarily chosen vertex from each $P_2$, and every vertex of degree~1 from each other tree.  Pair each vertex in $A$ with one of its neighbours in $\ell(F_S')$ (it must have at least one by the definition of $F_S'$).  These pairs induce a collection of pairwise non-adjacent $P_2$'s, and, as $G$ is $sP_2$-free, $|A| \leq s-1$.  Notice that from a vertex in $A$, it is not possible to find a path containing more than $3s-2$ vertices within $F_S'$ since a longer path contains $sP_2$ as an induced subgraph.
We choose a path of length at most $3s-2$ from each vertex~$v\in A$ to a fixed vertex of the component of $F_S'$ containing~$v$. In this way the union of all these paths cover all the vertices of $F_S'$, and the claim follows.

\medskip
\noindent
Let $J(F_S') \subseteq S$ be the set of vertices in $G-V(F_S')$ that have at most one neighbour in $F_S'$.  Notice that $\ell(F_S') \subseteq J(F_S')$.

\begin{claim}
$\ell(F_S')$ is a maximal independent set in $G[J(F_S')]$.
\end{claim}
To prove the claim, suppose instead that there is a vertex $v \in J(F_S') \setminus \ell(F_S')$ that has no neighbour in $\ell(F_S')$.  Then $G[V(F_S)\cup \{v\}]$ is a forest as $v$ has at most one neighbour in $F_S$, and so $S \setminus \{v\}$ is a feedback vertex set of $G$, contradicting the minimality of $S$ and proving the claim.

\medskip
\noindent
We can find minimal feedback vertex sets of $G$ as follows.  First we choose a
skeleton
on at most $3s^2-5s+2$ vertices, and then, from amongst those vertices that have at most one neighbour in the chosen
skeleton, we choose a maximal independent set.  Then the set of vertices in neither the forest nor the independent set is tested to see whether it forms a minimal feedback vertex set.   If we consider all possible choices of forest and independent set, then each minimal feedback set will be found as at some point a
corresponding
skeleton and its leave will be chosen.
By our first claim, there are
$O(n^{3s^2})$
choices for the forest, and, by Theorem~\ref{t-by}, there are $O(n^{2s})$ choices for the independent set.  The theorem is proved.
\qed
\end{proof}

We also need the following result on the price of connectivity for feedback vertex set due to Belmonte, van 't Hof, Kami\'nski, and Paulusma.
Notice that this result holds even for $sP_3$-free graphs. The constant $12s^2-2s-2$ is determined in the proof of Lemma~9 of~\cite{BHKP17}.

 \begin{theorem}[\cite{BHKP17}]\label{t-price}
Let $s\geq 1$ and let $G$ be a connected $sP_3$-free graph. Let $f$ be the size of a minimum feedback vertex set of
$G$. Then the size of a minimum connected feedback vertex set of $G$ is at most $f+12s^2-2s-2$.
\end{theorem}

We are now ready to prove our result on the complexities of {\sc Feedback Vertex Set} and {\sc Connected Feedback Vertex Set} restricted to $sP_2$-free graphs.

\begin{theorem}\label{t-main}
For every constant $s\geq 1$, {\sc Feedback Vertex Set} and {\sc Connected Feedback Vertex Set} can be solved in polynomial time for $sP_2$-free
graphs.
\end{theorem}

\begin{proof}
We will show that there are polynomial-time algorithms that find the minimum size sets.  The theorem, which concerns decision problems, follows.

Let $G$ be an $sP_2$-free graph. We may assume without loss of generality that $G$ is connected.
By the combination of Lemma~\ref{l-1} with Theorem~\ref{t-ss}, we can in fact enumerate all minimal feedback vertex sets of $G$ in polynomial time.
Thus the result for {\sc Feedback Vertex Set} follows.

For {\sc Connected Feedback Vertex Set}, we can apply Lemma~\ref{l-polyfind}.  We need only note that the property of being a feedback vertex set is inherited by supersets and that, by Theorem~\ref{t-price}, within the class of $sP_2$ graphs, the minimum size of a connected feedback vertex set differs from the size of a minimum size set by at most a constant.
\qed
\end{proof}

\subsection{Odd Cycle Transversal and Connected Odd Cycle Transversal}\label{s-o}

We now consider the {\sc Odd Cycle Transversal} problem and its connected variant.
Again we can use a result on the price of connectivity,
which is also covered by Lemma~10 of~\cite{HJMP16}.

\begin{theorem}[\cite{HJMP16}]\label{t-price3}
Let $s\geq 1$ and let $G$ be a connected $sP_3$-free graph. Let $f$ be the size of a minimum odd cycle transversal of
$G$. Then the size of a minimum connected odd cycle transversal of $G$ is at most $f+4s^2+2s-10$.
\end{theorem}

We can now prove our final result.

\begin{theorem}\label{t-main3}
For every constant~$s\geq 1$,  {\sc Odd Cycle Transversal} and {\sc Connected Odd Cycle Transversal} can be solved in polynomial time for $sP_2$-free graphs.
\end{theorem}

\begin{proof}
Let $S$ be a minimal odd cycle transversal of an $sP_2$-free graph $G$. Let $F_S=G-S$. Then $F_S$ is a bipartite graph.
We choose a bipartition $(X,Y)$
of $F_S$
such that $X$ has maximum size (so every vertex in $Y$ has at least one neighbour in $X$). Then $X$ is a maximal independent set of $G$, as otherwise there exists a vertex~$u\in S$ not adjacent to any vertex of $X$, and thus $S\setminus \{u\}$ is an odd cycle transversal of $G$, contradicting the minimality of $S$. Moreover, by a similar argument, $Y$ is a maximal independent set of~$G-X$.

We describe a procedure to find all minimal odd cycle transversals of $G$, and, as the minimum size transversals of $G$ will be amongst them this provides an algorithm for {\sc Odd Cycle Transversal}.  We enumerate all maximal independent sets of~$G$, and for each maximal independent set~$X$, we enumerate all maximal independent sets of~$G-X$.
For each such set $Y$, we note that $V(G)\setminus (X\cup Y)$ is an odd cycle transversal of $G$.  By the arguments above, we will find every minimal odd cycle transversal in this way, and, by Theorems~\ref{t-by} and~\ref{t-tias}, this takes polynomial time.

For {\sc Connected Odd Cycle Transversal}, we can now apply Theorem~\ref{t-price3}
and Lemma~\ref{l-polyfind} as the property of being an odd cycle transversal holds for supersets.
\qed

\end{proof}

\noindent
{\bf Remark 1.} Just as for minimal vertex covers, we can compute all minimal feedback vertex sets and all minimal odd cycle transversals of an $sP_2$-free graph in polynomial time for every constant $s\geq 1$.

\section{Conclusions}\label{s-con}

We first showed that just like the {\sc Connected Vertex Cover} problem,
the {\sc Connected Feedback Vertex Set} and {\sc Connected Odd Cycle Transversal}
problems are \NP-complete for $H$-free graphs if $H$ contains a cycle or a claw.
Hence, we could restrict ourselves to graphs $H$ that are linear forests. We then proved
that all three problems can be solved in polynomial time for $sP_2$-free graphs.

Our main goal was to show an application of the price of connectivity for cycle transversals
(Theorems~\ref{t-price2},~\ref{t-price}, and~\ref{t-price3}).
 In fact these three structural results on the price of connectivity hold even for $sP_3$-free graphs. Hence, a natural future research direction would be to try to extend our algorithmic results to $sP_3$-free graphs. However, then we can no longer bound the number of minimal transversals as before, so the situation is not clear.

It would also be interesting to research whether it is possible to compute, just as for the three original problems (see Remark~1), all minimal connected transversals in polynomial time for the three variants studied, even when we restrict
the input to $2P_2$-free graphs. With respect to this question, we recall an open problem of
Golovach, Heggernes, and Kratsch~\cite{GHK15}. They asked if there exists a graph class with a polynomial number of minimal vertex covers but an exponential number of minimal connected vertex covers.

\subsubsection*{Acknowledgement}
We are grateful to Andrea Munaro and an anonymous referee for helpful remarks.

\end{document}